# Structure of long-period stacking/order Mg-Zn-RE (RE: rare-earth and Y) phases with extended non-stoichiometry ranges

D. Egusa and E. Abe*

*Department of Materials Science & Engineering, University of Tokyo, Tokyo 113-8656, Japan*

*Corresponding author. Tel.: +81 3 5841 7167; fax: +81 3 5841 7167; e-mail: abe@material.t.u-tokyo.ac.jp

We propose structure models of the unique long-period stacking/order (LPSO) phases formed in the Mg-Zn-RE alloys, based on Z-contrast scanning transmission electron microscopy (STEM) observations and first-principles calculations. The LPSO structures are long-period stacking derivatives of the *hcp*-Mg structure, and the Zn/RE distributions are restricted at the four close-packed atomic layers forming local *fcc*-stacking (i.e., a local ABCA stacking). Chemical order is well developed for the LPSO phases formed in $Mg_{97}Zn_1Er_2$ (14H-type) and $Mg_{85}Zn_6Y_9$ (18R-type) alloys with pronounced superlattice reflections, and the relevant Zn/RE distributions are clearly emerged in the Z-contrast atomic images. Initial ternary-ordered models are constructed by placing all the atoms at the ideal honeycomb sites, leading to plausible space groups of $P6_3/mcm$ for 14H-type and $C2/m$, $P3_112$ or $P3_212$ for 18R-type. Characteristic ordered feature is well represented by the local $Zn_6RE_8$ clusters, which are embedded in the *fcc*-stacking layers in accordance with the $L1_2$-type short-range order. Energy-favored structural relaxations of the initial model cause significant displacements of the Zn/RE positions, implying that strong Zn-RE interactions may play a critical role for the phase stability. The LPSO phases seem to tolerate a considerable degree of disorder at the Zn and RE sites with statistical co-occupations by Mg, extending the non-stoichiometry phase region bounded along the Zn/RE equi-atomic line from ~$Mg_{94.0}Zn_{2.0}Y_{4.0}$ to ~$Mg_{83.3}Zn_{8.3}Y_{8.3}$.

## 1. Introduction

Mg alloys are an important light-weight structural material [1] with their remarkable specific strength, i.e., strength-to-weight ratio. On this regard, dilute magnesium alloys containing a few atomic percent of Zn/RE (RE denotes rare-earth (RE) and Y atoms) have attracted increasing attentions, because many of the Mg-1at.%Zn-2at.%RE (denoted as $Mg_{97}Zn_1RE_2$ hereafter) alloys are shown to reveal the strength higher than ~350MPa by simply applying a conventional hot-extrusion process for the as-cast ingots [2]. In particular, the strength can be maximized up to ~600MPa when a non-equilibrium, rapid solidification processing is applied for the $Mg_{97}Zn_1Y_2$ alloy [3], realizing the specific strength over 300MPa g/cm$^3$ that is even beyond that of Al alloys (e.g., extra-super-duralumin) or Ti alloys (e.g., Ti-6at.%Al-4at.%V). The key microstructures of these Mg-Zn-RE alloys are a novel type of long-period structure phases [4-8], which are believed to play a critical role to realize the superior mechanical properties. The novel structures are long-period chemical-ordered as well as stacking-ordered [5]; a unique long-period stacking/order (LPSO) structure (this wording follows the sense that, in metallurgy, the "order" specifically means a chemical order, which is generally termed for configurational ordering of different atom species on the fixed atomic sites). Interestingly, the apparent chemical order occurs with confined Zn/RE distributions so as to synchronize with the relevant stacking order [9-12], and hence the Mg-Zn-RE LPSO structures are referred to as the synchronized LPSO [12]. Because of the restricted Zn/Y distributions, the LPSO phases are able to form with relatively less Zn/RE contents, enabling to gain a considerable amount of their volume even in the dilute Mg-Zn-RE alloys [2, 12]. This is important for a design of the (LPSO + α-Mg) two-phase practical alloys [2], whose mechanical properties can be tuned by a phase-volume ratio along with a concept of composite materials [13-15].

The Mg-Zn-RE LPSO structures are fundamentally long-period stacking variants of a hexagonal-close-packed (*hcp*) structure of the Mg crystal, and accordingly rhombohedral (R) and hexagonal (H) Bravais lattices alternatively appear depending on the stacking period of the close-packed atomic layers. So far, four polytypes, 10H, 18R, 14H, and 24R have been reported for the Mg-Zn-RE alloys [8, 12]. Z-contrast scanning transmission electron microscope (STEM) observations of the LPSO phases clearly showed up a chemical order along the stacking direction [9-12], as displayed by significant Zn/RE enrichments at the particular two close-packed layers relevant to faulting of the original 2H-type stacking of the *hcp* structure. Consequently, it was found that the LPSO polytypes are systematically described by the common structural unit composed of local AB'C'A stacking, where C represents the fault layer with respect to the original AB stacking of *hcp*, and B' and C' denote the layers enriched by Zn and RE atoms [9-12]. Here, it is noted that the ABCA stacking-unit defines a local face-centered cubic (*fcc*) stacking. The chemical order of the LPSO phases is further developed within the Zn/RE enriched close-packed planes, as monitored by pronounced superlattice reflections in the relevant diffraction pattern, even though the degree of order seem to vary significantly depending on the phase compositions.

That is, the superlattice peaks appear distinctly for the LPSO phases with rich Zn/RE contents, e.g., ~$Mg_{85}Zn_7Y_7$ [16] or ~$Mg_{84}Zn_8Y_8$ [11], whereas they tend to become significantly weak (almost disappear) for the phases with less Zn/RE contents, e.g., ~$Mg_{93}Zn_2Y_4$ [5, 12, 17] or ~$Mg_{92}Zn_2Y_6$ [18]; detailed features will be described later with Fig. 2. For the well-ordered Zn/Y-rich LPSO phases of 18R-type and 14H-type, possible ordered arrangements within the close-packed planes have been proposed recently [11]. In the model structures, however, Zn/Y atoms are distributed in a significantly anisotropic manner with only two-fold rotation symmetry in the [0001] projection (the Zn/Y distributions are only shown for the monolayer in Ref. 11), and the complicated variant configurations are forced to satisfy the apparent six-fold (or three-fold) symmetry of the LPSO structures. These features were originated not from confident experimental evidences such as Z-contrast imaging, but rather from simple trials of placing Zn/Y atoms at the honeycomb sites, for which neither the space group symmetry nor atom site symmetry were assumed during the model constructions. Therefore, we would presume there still remain ambiguities in terms of the Zn/RE ordered arrangements within the LPSO structures.

In the present work, we investigate details of the LPSO structures based on atomic-resolution Z-contrast imaging along the several major zone axes, and construct the model structures with plausible space group symmetries. We further attempt to tune the models along energy-favored relaxations based on first-principles calculations, aiming as well to deepen the understanding of an atomistic origin of structural stabilities. As described in the present paper, the ternary LPSO structures with distinct assignments of Mg/Zn/RE sites have been successfully constructed. In the present ideal stoichiometry LPSO models, the RE sites are not only limited to the two close-packed layers, as assumed in the previous works [9-12], but are extended over the four layers; the RE atoms are located at all the layers in the ABCA-unit of the local *fcc*-stacking. Very recently, this feature has been also reported for the LPSO phase in a Mg-Al-Gd alloy [19]. Therefore, the apparent two-layer Zn/RE enrichment in the previous Z-contrast observations is reasonably attributed to substitution disorder at the Zn and/or RE sites that are able to tolerate mixed occupations with Mg, allowing non-stoichiometry formation of the LPSO phases even with the significantly dilute Zn/RE contents. This also explains why the apparent degree of order significantly depends on the LPSO compositions, as will be discussed in the present paper.

## 2. Experimental procedure

Nominal compositions of the Mg-Zn-RE alloys used in the present study are $Mg_{97}Zn_1Y_2$, $Mg_{97}Zn_1Er_2$ and $Mg_{85}Zn_6Y_9$ (at.%). Master alloy ingots were prepared either by high frequency induction melting of pure metals in an argon atmosphere, or by furnace melting pure metals in a steel tube under a $CO_2$ atmosphere, followed by casting into a water-cooled Cu mold (see Ref. 2 and Ref. 20 for further details of sample preparations). In order to obtain the less-ordered LPSO phases, the $Mg_{97}Zn_1Y_2$ alloy ingots were rapidly solidified (RS) by a single-roller melt-spinning method with wheel rotating speed of approximately 42 m/s, and the RS-ribbon specimens were annealed at 673K for 48h [21]. The well-ordered LPSO phases were successfully developed for the $Mg_{97}Zn_1Er_2$ alloy annealed at 573K for 168h and the $Mg_{85}Zn_6Y_9$ alloy annealed at 673K for 72h. All the heat treatments were performed using a Pyrex tube in an argon atmosphere after evacuation to pressures lower than $3\times10^{-3}$ Pa, followed by air-cooling down to the room temperature. The LPSO phases coexist with the α-Mg phase in the annealed $Mg_{97}Zn_1Y_2$ and $Mg_{97}Zn_1Er_2$ alloys, whereas the LPSO phase (18R-type) is obtained as an almost single-phase for the annealed $Mg_{85}Zn_6Y_9$ alloy; see Fig. 1. As confirmed by a transmission electron microscope (TEM) observation, the microstructure consists of equiaxed LPSO grains with a volume fraction larger than ~95%, coexisting with occasional particles of the $Mg_{24}Y_5$ compound and the secondary phase (unidentified) finely distributed at the grain boundaries. Therefore, the nominal composition $Mg_{85}Zn_6Y_9$ is approximately relevant to the LPSO phase composition.

Thin foils for TEM/STEM observations were prepared by standard Ar-ion milling. For atomic-resolution high-angle annular dark-field (HAADF)-STEM observations, we used a conventional 200 kV STEM (JEM-2010F) and an aberration-corrected 200 kV STEM (JEM-ARM200F) that provides a minimum probe of approximately ~0.09 nm with a convergence semi-angle α ~25mrad. For HAADF imaging, the annular detector was set to collect the electrons scattered at angles between 70-150mrad sufficiently high to reveal a chemical sensitive Z-contrast. Image simulations are carried out using the WinHREM software based on the fast-Fourier-transform multislice algorithm [22], and the foil thicknesses used for calculations were estimated according to the log-ratio method by electron energy loss spectroscopy.

## 3. Experimental results

### *3.1. Degree of order of the LPSO phases*

Figs. 2(a)-(i) show electron diffraction patterns taken from the several zone axes of the LPSO phases, which are formed in the $Mg_{97}Zn_1Y_2$ ((a)-(c)), $Mg_{97}Zn_1Er_2$ ((d)-(f)) and $Mg_{85}Zn_6Y_9$ ((g)-(i)) alloys. The indices are represented in accordance with the *hcp* structure, such as $[\bar{1}2\bar{1}0]_{hcp}$ for the H-type LPSO structures ((a)-(f)), while they are given based on the fundamental rhombohedral lattice for the R-type LPSO structure ((g)-(i)). By using energy-dispersive x-ray spectroscopy with TEM, the compositions of the LPSO phases precipitated in the $Mg_{97}Zn_1Y_2$ and $Mg_{97}Zn_1Er_2$ alloys are estimated to be approximately Mg-2.0±1.0at.%Zn-4.0±2.0at.%Y and Mg-11.0±1.0 at.%Zn-6.0±2.0at%Er, respectively (denoted as $Mg_{94}Zn_2Y_4$ and $Mg_{83}Zn_{11}Er_6$ hereafter). For the $Mg_{85}Zn_6Y_9$ alloy, the nominal composition can be relevant to that of the LPSO phase (almost single-phase formation: Fig. 1). Therefore, for the $Mg_{97}Zn_1Er_2$ and $Mg_{85}Zn_6Y_9$ alloys, the LPSO phases occur with sufficient Zn/RE contents comparable with that of the previously reported well-ordered LPSO phases [11, 16]. On the other hand, the LPSO phase forms with significantly less Zn/RE contents for the $Mg_{97}Zn_1Y_2$ alloy [5, 7, 12, 18]. From the diffraction peak appearances along the *c**-directions, the fundamental stacking structures are indentified as 14H-type for the $Mg_{94}Zn_2Y_4$ and the $Mg_{83}Zn_{11}Er_6$ precipitated phases, and as 18R-type for the $Mg_{85}Zn_6Y_9$

phase; hereafter, these are denoted as 14H-$Mg_{94}Zn_2Y_4$, 14H-$Mg_{83}Zn_{11}Er_6$ and 18R-$Mg_{85}Zn_6Y_9$.

A degree of order of the LPSO phase is found to depend significantly on the compositions, as clearly seen in the series of diffraction patterns. In the patterns taken with the incident beam along the stacking direction, a large number of weak peaks distinctly appear for the 14H-$Mg_{83}Zn_{11}Er_6$ (Fig. 2 (d)) and 18R-$Mg_{85}Zn_6Y_9$ (Fig. 2 (g)), while they are extremely weak for the 14H-$Mg_{94}Zn_2Y_4$ (Fig. 2 (a)). These extra peaks are direct sign of the extended Zn/RE order within the close-packed planes, and their diffused appearances indicate that the order is not well developed in the long-range. Here, it should be remembered that the incidence normal to the close-packed plane can be a low-index crystallographic zone axis only for the hexagonal lattice, but not for the rhombohedral lattice; i.e., the diffraction peaks in Fig. 2(g) are not all exactly at the Bragg condition. Nevertheless, in the patterns of Figs. 2 (d) and (g), the extra peaks appear to be almost identical arrangements that represent approximately six-times period along $\langle\bar{1}2\bar{1}0\rangle_{hcp}$ directions, suggesting that similar local Zn/RE order occurs both in the 14H- and 18R-LPSO phases. In the patterns with the incidence normal to the stacking axis, these extra diffuse peaks are significantly streaked along the *c**-direction as indicated by arrowheads, and their intensity certainly increases along 14H-$Mg_{94}Zn_2Y_4$ (Fig. 2 (b), (c)) < 14H-$Mg_{83}Zn_{11}Er_6$ (Fig. 2 (e), (f)) < 18R-$Mg_{85}Zn_6Y_9$ (Fig. 2 (h), (i)). On this basis, the degree of order tends to be well developed for the Zn/RE-rich LPSO phases, particularly for the 18R-$Mg_{85}Zn_6Y_9$ that even reveals the sharp Bragg-like maxima within the diffuse peaks, as observed around $1/2\,(10\bar{1}\bar{1})_{hcp}$ (Fig. 2(h)) and $1/6\,(\bar{1}2\bar{1}0)_{hcp}$ (Fig. 2(i)). Here, it should be noteworthy that the splitting-maxima in Fig. 2(h) are found to occur at incommensurate positions. With this fact in mind, we further find that all the weak peaks are not exactly at the commensurate positions related to the $6\times(10\bar{1}0)_{hcp}$ order. Therefore, in a strict sense, the present Mg-Zn-RE LPSO phases are incommensurately modulated to some degree; though, we presently assume the $6\times(10\bar{1}0)_{hcp}$ commensurate superstructure for the purpose of constructing the average, ideal LPSO models. Details of the incommensurate features will be described elsewhere.

*3.2. Z-contrast imaging of the LPSO phases*

Fig. 3 shows Z-contrast STEM images of the 14H-$Mg_{94}Zn_2Y_4$ and 18R-$Mg_{85}Zn_6Y_9$, revealing the Zn/Y distributions as well as the relevant stacking structures. As reported previously [9-12], the LPSO structures essentially accompany the significant Zn/Y enrichment at the specific two close-packed layers relevant to faulting of the original 2H-type *hcp* structure, the feature of which was represented by the structural unit composed of local AB'C'A stacking [12]. In the present Z-contrast images, significant bright contrast representing Zn/Y enrichment indeed occurs at the corresponding two layers, but it is evidently not limited to these layers; see in particular the image of the 18R-$Mg_{85}Zn_6Y_9$ (Fig. 3(b)). As confirmed by the corresponding Z-contrast intensity profile, secondary-maxima certainly occur at the layers neighboring to the strongest two layers, as indicated by arrows, manifesting that the Zn and/or Y occupations are extended into the four close-packed layers. Accordingly, slight increase at the relevant layers in the 14H-$Mg_{94}Zn_2Y_4$ (Fig. 3(a)) should also be accounted for the intrinsic Zn/RE occupations with certain amounts, even though such small secondary-maxima were previously discussed in terms of possible artifacts of STEM imaging (sub-peak effects of the STEM probe shape [5]). On the basis of these facts, the intensity-profile differences between the 14H-$Mg_{94}Zn_2Y_4$ and the 18R-$Mg_{85}Zn_6Y_9$ can be reasonably related to the different Zn/RE contents within the extended four layers, and the Zn/RE occupations at the relevant atomic sites directly affect the degree of order of the LPSO phases. In fact, image simulations based on the present LPSO models, whose details will be described later, reproduce fairly well the observed Z-contrast, as inserted in each of the image by red frames (the compositions of each simulated LPSO structures are $Mg_{94.0}Zn_{2.0}Y_{4.0}$ and the 18R-stoichiometry ($Mg_{80.6}Zn_{8.3}RE_{11.1}$), as their details will be described later in the "Discussion" section. Parameters used for simulations are denoted in Fig. 3 caption). A good match between the experiments and simulations can be further confirmed by the layer-by-layer intensity distributions shown in the bottom of Fig. 3, even though there still remain slight discrepancies in the detailed profiles including background level (the discrepancies could be mostly due to intrinsic disorder in the present LPSO structures, which were hardly incorporated in the present simulations; nevertheless, the simulation profiles in Fig. 3 were one of the best-fit to reproduce the observed layer-by-layer Z-contrast intensity).

Details of the Zn/RE site distributions within the four close-packed planes can be investigated by the Z-contrast imaging along the several major zone axes of the LPSO structures. Before describing the characteristic Zn/RE order, it should be mentioned that, for the present 18R-$Mg_{85}Zn_6Y_9$ and 14H-$Mg_{83}Zn_{11}Er_6$ phases, the chemical order is found to be not developed into a long-range but limited to only a short-range; this seems to be consistent with the diffused appearances of the relevant superlattice peaks in Figs. 2(d) and (g). Even with this short-range order condition, the local Zn/RE configurations that are responsible for the $6\times(10\bar{1}0)_{hcp}$ chemical order can be successfully inferred directly from the Z-contrast images shown in Fig. 4, as described below.

In the images of the 18R-$Mg_{85}Zn_6Y_9$ taken along $[\bar{1}2\bar{1}0]_{hcp}$ and $[10\bar{1}0]_{hcp}$ directions, the ordered domains traced by local periodic arrays of the bright-spot are frequently observed, revealing the modulation period with every second spots along $[10\bar{1}0]_{hcp}$ direction (Fig. 4(a)) and every sixth spots along $[\bar{1}2\bar{1}0]_{hcp}$ direction (Fig. 4(b)), as indicated by dashed lines. These periods are consistent with the (average) ordered period estimated from the corresponding diffraction patterns (Fig. 2), and the brightest spots represent the atomic columns containing RE sites. Note that the brighter spots indicating the RE sites indeed occur at the extended layers next to the brightest ones, as indicated by red arrowheads in Figs. 4(a) and (b); for the latter, this is evidently recognized with characteristic local brightest-spot arrangements similar to 'ϕ' shape, although they appear to be not always completed due to overlapped domains along the incident beam direction. The RE site distributions can also be observed in the projection along the stacking direction; though, again, the incidence normal to the close-packed plane cannot be a low-index zone axis for the 18R-$Mg_{85}Zn_6Y_9$ with a rhombohedral lattice. Alternatively, Z-contrast image was obtained from the ordered 14H-$Mg_{83}Zn_{11}Er_6$ with the incidence along [0001] direction of a hexagonal lattice, as shown in Fig. 4(c). In the image, characteristic local contrasts composed of hexagonal arrays of the bright dots (dashed lines) are emerged, each of which accompanies the brightest dot at its center as indicated by red arrowheads. At some regions, local correlations between the brightest-dot arrangements appear to be well consistent with the superlattice order of $6\times(10\bar{1}0)_{hcp}$. However, it should be mentioned again that no long-range order was observed for the brightest-dot distributions in the [0001] projection, and their correlations are just limited in the short-range. As described later, we find that these brightest spots are originated from the two Er atoms projected along [0001] projection; i.e., all the bright spots indicated by red arrowheads in Figs. 4 (a)-(c) are due to the equivalent RE sites.

On the basis of above Z-contrast characteristics, for both the 14H and 18R LPSO we have successfully constructed ideal model structures with the Zn/RE ordered arrangements across the four close-packed layers, details of which are described in the next section. Simulated images based on the present LPSO models (the initial model in Table 1), inserted in the images Figs. 4(a)-(c), fairly well reproduce the experimental images when the RE atoms are located not limited to the two close-packed layers but over the four close-packed layers. Furthermore, the simulations based on energetically-relaxed structures (the energetically optimized model in Table 1) qualitatively reproduce the characteristic local features of the each bright spot (Figs. 4(a)-(c)), some of which are certainly diffused and/or elongated to some degree.

## 4. LPSO model structures

### *4.1. Average structural models*

Table 1 summarizes the 14H and 18R LPSO model structures with their unit-cell dimensions and atomic coordinates. During the model constructions, we assume appropriate space groups of $P6_3/mcm$ for 14H-type, and $C2/m$ (monoclinic) and $P3_212$ (trigonal) for 18R-type; these are all able to incorporate the identical Zn/RE order configurations at the common *fcc*-stacking layers. Note that the space group of $P3_112$ of a trigonal lattice is also a possible candidate for the 18R-LPSO, and the choice either monoclinic ($C2/m$) or trigonal ($P3_112$ or $P3_212$) depends on how to arrange the Zn/RE order along the stacking direction, as discussed later.

Firstly, we describe structural characteristics of the 14H-LPSO models. Ternary-ordered structure is initially constructed by placing Zn and RE atoms at the ideal honeycomb lattice sites (denoted by fractional numbers in Table 1), and the successful Zn/RE assignments are shown by the projection along the major zone axes in Figs. 5(a) and (b), which are comparable with the observed characteristics in the Z-contrast atomic images. It is well recognized that the RE sites are distributed across the four close-packed layers forming the local *fcc*-stacking (B'C'A'B' in Fig. 5(b)), which is used as the common structural unit of the LPSO polytypes [9-12]. The isolated RE sites (Er2 at 4*e* in Table 1) described earlier with Fig. 4 indeed occur at the *h*-layers adjacent to the *c*-layers, as indicated by red arrowheads in Fig. 5(b). It should be noted here that, as known from Jagodzinski notation, only the two *c*-layers are truly at the local *fcc*-environment within the *fcc*-stacking four layers; in this sense, the Zn/RE locations occur at the extended *fcc*-environment along the local *hcch* stacking (Fig. 5(b)). In the [0001] projection (Fig. 5(a)), the local hexagonal arrays of RE atoms (dashed hexagon) accompany the inner Zn atoms as well as the centered RE atoms, a network of which represents the superlattice order of $6\times(10\bar{1}0)_{hcp}$ as indicated by rhombus (hexagonal lattice setting of the 14H is defined as orthogonal to that of *hcp*-Mg, as shown in bottom-left in Fig. 5(a)). By looking at the layer-by-layer Zn-RE configurations in Fig. 5(c), it is noticed that the RE atoms are arranged in accordance with the $D0_{19}$-type order, which is an ordered version of *hcp* and known to occur frequently for the Mg-RE alloys [1, 23, 24]. All the Zn atoms are located at the nearest-neighbor positions to the RE sites including those at the different layers, forming the local Zn-RE order of $L1_2$-type that is an ordered version of *fcc*, as shown later in Figs. 7 (a)-(c). Therefore, the long-range chemical order feature of the LPSO is well represented by the arrangements of the $L1_2$-type $Zn_6RE_8$ clusters embedded in the *fcc*-stacking layers. Occurrence of such Zn-RE clusters implies strong Zn-RE interactions, as discussed in the next section based on first-principles calculations.

Having the above characteristic order represented by the $L1_2$-type $Zn_6RE_8$ clusters as motifs, the 18R-LPSO model structure is constructed by considering the locations of the $Zn_6RE_8$ clusters in the 18R-type stacking structure. Within the local *fcc*-stacking layers, it is reasonable to assume the $Zn_6RE_8$-cluster arrangements identical with that of the 14H-type (Fig. 5(a); $6\times(10\bar{1}0)_{hcp}$ order), and therefore the 18R-LPSO model depends on how to arrange the $Zn_6RE_8$ clusters along the stacking direction, i.e., inter-layer order. There are basically two ways for arranging the $Zn_6RE_8$ clusters to be reasonably incorporated in the 18R-stacking composed of the three *fcc*-stacking units, leading to the possible space groups of $C2/m$ (monoclinic) or $P3_112$ or $P3_212$ (trigonal) as the ordered 18R, as shown in Fig. 6. Note that, in the projection along $[\bar{1}2\bar{1}0]_{hcp}$ (Fig. 6(a)), these structures appear to be identical Zn/RE ordered distributions, while their inter-layer order differences can be clearly distinguished in the $[10\bar{1}0]_{hcp}$ projection in Figs. 6(b) and (c). The monoclinic structure is obtained when the $Zn_6RE_8$ clusters are arranged in accordance with the successive displacements along the one fixed direction of $[10\bar{1}0]_{hcp}$, as shown by the arrows in Fig. 6(d) (displacement magnitude is $2\times(10\bar{1}0)_{hcp}$), and the resultant inter-layer order between the $Zn_6RE_8$ clusters are viewed as XXX in $[10\bar{1}0]_{hcp}$ projection and XYZ in $[01\bar{1}0]_{hcp}$ projection, as shown in Fig. 6(b). On the other hand, the trigonal structure results from the cyclic displacements along with the three equivalent $\langle 10\bar{1}0\rangle_{hcp}$ directions, as also indicated by the arrows in Fig. 6(e), and the inter-layer arrangement of the $Zn_6RE_8$ clusters is represented as XXY along the stacking direction, as shown in Fig. 6(c). Here we note in particular that, for the present 18R-$Mg_{85}Zn_6Y_9$ phase with significantly diffused streaks along the $c$*-directions (Fig. 2(i)), it is impossible to uniquely identify neither the likely space group nor the one of these structures. In fact, no long-range inter-layer order was observed for the arrangement of the $Zn_6RE_8$ clusters along the stacking direction; only the short-range correlation was observed, as described before with Figs. 4(b) and (c). Therefore, the present LPSO phases may well be viewed as a consequence of the short-range inter-cluster order, which is defined by $6\times(10\bar{1}0)_{hcp}$ transverse correlation within the *fcc*-stacking units and the relative $2\times(10\bar{1}0)_{hcp}$ displacements for the inter-layer correlation along the stacking direction. At this stage, we do not have any clear ideas for the microscopic origin of such cluster-cluster interactions; nevertheless, it should be emphasized again that the above inter-cluster correlations are essentially necessary to be consistent with the distinct weak peaks in the diffraction patterns (e.g., suppose that the clusters are stacked randomly by $(10\bar{1}0)_{hcp}$ displacements instead of $2\times(10\bar{1}0)_{hcp}$ displacements, no superlattice peaks would appear in the patterns of the close-packed planes of Figs. 2(d) and (g)). It is noteworthy here that, even for the very well ordered 18R-LPSO phase recently identified in the Mg-Al-Gd alloy [19], the long-range inter-layer order is not realized; the $A_6B_8$-type clusters stack in a disordered manner but nevertheless being restricted by the $2\times(10\bar{1}0)_{hcp}$ displacements along one of the three directions denoted in Fig. 6(e). Despite of this weak inter-layer interaction between the local Zn-RE clusters, it is worthwhile to remember that the stacking periodicity always appears to be robust long-range order, indicating that composition/cluster fluctuations within the layers do not much affect the fundamental long-period stacking structures. This is supported by the fact that the compositions of the Zn/RE-enrich layers are found to be identical for the LPSO polytypes with different periodicity [12]. In this sense, the LPSO structures may well be a consequence of primary ordering of the fundamental stacking structures, followed by chemical ordering favored at the local *fcc*-layers that seem to be essential to incorporate Zn/RE atoms simultaneously within the structure.

*4.2. Structure tuning based on first-principles calculations*

The initial model structures with the successful Zn/RE assignments at the original honeycomb atomic sites, as described with Figs. 5 and 6, are further tuned based on first-principles calculations [25, 26]. Calculations were performed for the LPSO structures with ideal stoichiometry, 14H-$Mg_{35}Zn_3Er_4$ and 18R-$Mg_{29}Zn_3Y_4$ (monoclinic and trigonal) by using the Vienna Ab initio Simulation Package (VASP) within the framework of density-functional theory (DFT) [27], based on generalized gradient approximation (GGA) [28] and ultrasoft scalar relativistic pseudo-potentials (detailed conditions/parameters are described in Fig. 7 caption). After the energetically favored structural optimizations, the energy differences gained from the initial configurations were -140.2 kJ/mol and -66.7 kJ/mol for the 14H-$Mg_{35}Zn_3Er_4$ and 18R-$Mg_{29}Zn_3Y_4$, respectively. During the optimizations, there were almost no significant changes in lattice parameters (details are described in Table 1 caption), and the optimized atomic coordinates are denoted by decimal fractions in Table 1. By comparing the local atomic configurations before and after the structural relaxations, it is found that significant displacements commonly occur for the Zn and RE atom positions, behaviors of which are exemplified for the 14H-$Mg_{35}Zn_3Er_4$ in Fig. 7. As described earlier, the Zn/Er atoms are arranged in accordance with the $L1_2$-type order within the local *fcc*-stacking layers (Fig. 7(a)-(c)), and the original *fcc*-based configurations are remarkably distorted after the relaxation, as shown in Figs. 7(d)-(f). With respect to the center of the original cubic, Zn atoms shift outward (about 0.90 Å) from it, while Er atoms move toward (about 0.58 Å) to it, resulting in the shorter Zn-Er distances from ~3.2 Å to ~3.0 Å. This indicates the strong interactions between Zn and Er atoms and seems to be consistent with large negative values of their mixing enthalpy $\Delta H_{mix}$, which is believed to be one of the essential empirical factors for the LPSO formation [2]. Such condensations of the Zn-RE atoms cause a puckering of the close-packed layers, as seen in Fig. 7(f), the feature of which can be related to elongated/diffused appearances of the relevant spots in the atomic images of Figs. 4(a) and (b). In fact, these local features are qualitatively reproduced by the image simulations, as described before. On the other hand, the Mg atoms around the Zn/RE atoms are transversely displaced within the close-packed planes about an order of ~0.1Å, revealing modulations such as seen at the edges of the projected structure in Fig. 7(e). These local relaxation behaviors around the $Zn_6RE_8$ clusters are essentially equivalent with those occurred for the 18R-$Mg_{29}Zn_3Y_4$, even though the displacement magnitudes were slightly different. We have further confirmed that the almost equivalent relaxations take place for the LPSO structures by switching Y-Er atoms; i.e., 14H-$Mg_{35}Zn_3Y_4$ and the 18R-$Mg_{29}Zn_3Er_4$.

A validity of the optimized LPSO structures is further confirmed by simulations of electron diffraction patterns. As evidently seen in Fig. 8, characteristic intensity distributions of the weak reflections in the experimental pattern (Fig. 8(a)), such as indicated by arrows, are reproduced not in the simulated pattern of the initial model (Fig. 8(b)) but fairly well in the pattern of the structure tuned by first-principles calculations (Fig. 8(c)). This strongly supports the fact that the Zn/RE order within the close-packed planes indeed accompanies significant displacive modulations, as represented in Figs. 7(d)-(f).

## 5. Discussion

### *5.1. Non-stoichiometry ranges of the LPSO phases*

With a successful construction of the ideal LPSO model structures, we are now able to derive their stoichiometry compositions as described in the previous section, 14H-$Mg_{35}Zn_3RE_4$ ($Mg_{83.3}Zn_{7.2}RE_{9.5}$) and the 18R-$Mg_{29}Zn_3RE_4$ ($Mg_{80.6}Zn_{8.3}RE_{11.1}$), which should be compared with the practical LPSO compositions. Figure 9 summarizes the LPSO compositions that have been experimentally determined so far, and their distributions are plotted in a schematic ternary phase diagram near the Mg corner. Since the heat treatment conditions were different for the each LPSO phase, the diagram of Fig. 9 is not an exact isothermal section but ranged over temperatures 573K ~ 793K, at which the LPSO phases indeed occur as highly thermodynamically stable conditions. By looking at the composition distributions with respect to the stoichiometry, it is immediately noticed that all the reported LPSO phases form with non-stoichiometry compositions extended to a less Zn/RE side. When the LPSO phases are formed in the $Mg_{97}Zn_1Y_2$ alloys including those prepared via the rapid-solidification processes, their compositions are commonly located at very dilute regions as shown by ①~④ in Fig. 9, and the relevant LPSO phases essentially show up with the extremely low-degree of order (Fig. 2(a)-(c)). On the other hand, the LPSO phases with relatively Zn/RE-rich conditions closer to the stoichiometry, as distributed in crowd ⑤~⑧ in Fig. 9, are indeed found to be well-ordered with distinct superlattice reflections (Fig. 2(g)-(i)). On these bases, the degree of order of the LPSO phases depends on the occupation conditions of the Zn/RE sites, which are able to tolerate a considerable degree of substitution disorders that allows the extended non-stoichiometry phase formation. It is noteworthy here that, even by annealing the present LPSO specimens at several high temperatures up to ~773K, we did not observe any order-disorder transformation represented by appearance-disappearance of the weak reflections. In addition, no peak-sharpening was observed for the diffused weak reflections during the further annealing, indicating that the long-range ordered LPSO phases are hardly realized for the present LPSO specimens.

### *5.2. Local chemical disorder in the LPSO phases*

Local occupation behaviors of the Zn/RE atoms in the LPSO phases are effectively discussed in terms of the $Zn_6RE_8$ clusters, which characterize well the chemical order features of the present LPSO structures. It is an important aspect that, for the stoichiometry LPSO phase, the density of the $Zn_6RE_8$ clusters in the *fcc*-stacking layers are already maximized within the framework of $6\times(10\bar{1}0)_{hcp}$ order; see again Fig. 5. This in turn means that the LPSO phases may hardly incorporate the extra $Zn_6RE_8$ clusters beyond the stoichiometry. Therefore, if it is difficult for the extra Zn and/or RE atoms beyond stoichiometry to solute into the *hcp*-stacking as well as *fcc*-stacking layers with the maximized $Zn_6RE_8$ clusters, the maximum tolerations of the Zn/RE atoms into the LPSO phases are strongly restricted by their stoichiometry compositions. This qualitatively explains why the non-stoichiometry phase-region appears not to extend towards the Zn/RE-rich side beyond the stoichiometry (Fig. 9). It should be mentioned that the 18R-LPSO phase recently reported for the Mg-Al-Gd alloy occurs at the composition very close to the relevant stoichiometry, and hence the cluster density seems to be almost maximized as confirmed by STEM observation [19].

Occurrences of the LPSO phases at the significantly dilute compositions may also be related to formation behaviors of the $Zn_6RE_8$ clusters. It is interesting to note that the non-stoichiometry ranges of the Mg-Zn-Y LPSO phases reveal anisotropic extensions that seem to be bounded by a fixed Y/Zn ratio around ~1, which is close to the RE/Zn ratio of the $Zn_6RE_8$ clusters (4/3) as indicated by the dashed line in Fig. 9. This fact supports well the view that the LPSO formation is primary governed by the $Zn_6RE_8$-cluster formation; i.e., the local $Zn_6RE_8$ configurations are basically robust, and their density/distributions within the *fcc*-stacking layers determines the LPSO compositions as well as the relevant degree of order. However, even with such cluster-constituent criteria, the local $Zn_6RE_8$ clusters may not be always completed for the non-stoichiometry LPSO phases with less Zn/RE contents, and they need to incorporate somehow partial substitutions by Mg atoms. Considering the atomic size, RE atoms (atomic radius $R$ : $R_{Y}$ ~ 1.80Å, $R_{Er}$ ~ 1.76Å) are likely to be replaced by the Mg atoms ($R_{Mg}$ ~ 1.60Å). On the basis of first-principles calculations, it is confirmed that the substitution for the isolated RE site at the *h*-layers (Fig. 5(b), Fig. 6(a)) is slightly energetically favored rather than that for the RE site at the *c*-layers; e.g., the total energy of the 18R-$Mg_{117}Zn_{12}Y_{15}$ (*C*2/m), being one Y atom replaced by the Mg atom, differs approximately 13.8 kJ/mol between the Y2-site (4*i* at the *h*-layers) and Y3-site (4*i* at the *c*-layers) substituted cases. Accordingly, it is presumed that the $Zn_6RE_6(Mg_2)$ clusters lacking the original two RE atoms at the *h*-layers are frequent configurations in the non-stoichiometry less-Zn/RE LPSO phases, accounting well for the phase occurrence around the Zn/Y ratio ~1 with the composition ~$Mg_{83.3}Zn_{8.3}Y_{8.3}$ (18R-$Mg_{30}Zn_3Y_3$). It is noteworthy in particular that the preferential Mg substitutions for the RE sites at the *h*-layers are consistent with the fact that the dilute LPSO phase reveals the significant Z-contrast only at the two close-packed layers (Fig. 3(a)); remember that the local *fcc*-stacking is constructed by *hcch* layers (Fig. 5(b), Fig. 6(a)). For further substitutions to be consistent with the very dilute LPSO phases, such as shown ①~③ in Fig. 9, both the Zn and RE sites are heavily replaced with the Mg atoms by more than ~60%, resulting in the extremely weak appearances of the superlattice peaks (Fig. 2(a)). During such a large amount of substitutions, it may be worthwhile to consider that the simultaneous replacement of Zn and RE atoms by the two Mg atoms (i.e., Zn-RE pair atoms ↔ 2×Mg atoms) would minimize the resultant local strains, since the average atomic size of Zn and RE becomes much closer to that of the Mg atom; e.g., average atomic radius between Zn ($R_{Zn}$ ~ 1.34Å) and Y ($R_Y$ ~ 1.80Å) is ~ 1.57Å, which is very close to the Mg atomic radius ~1.60Å. This scenario explains why the LPSO phases still occur around the Zn/Y ratio ~1 even at the very dilute compositions (Fig. 9), and seems to be well supported by the fact that the Zn-Y still forms the clusters (i.e., at the nearest-neighbor configurations) even in the very dilute LPSO phases, as confirmed by extended X-ray absorption fine structure (EXAFS) analysis [29]. Of course, such substitution behaviors involving the simultaneous removals of Zn-RE atoms by a pair may slightly differ depending on the degree of the Zn-RE interactions. In fact, the LPSO phase in the Mg-Zn-Er system appears at the composition deviated from the LPSO range of the Mg-Zn-Y system. Further precise and systematic structural analysis, including EXAFS to investigate local environments around the Zn or RE atoms, will deepen the understanding why the LPSO phases are able to form at dilute compositions far from the stoichiometry.

## 5. Conclusions

We have investigated the structures of the LPSO phases formed in $Mg_{97}Zn_1Y_2$, $Mg_{97}Zn_1Er_2$ and $Mg_{85}Zn_6Y_9$ alloys, based on electron diffraction and Z-contrast STEM experiments. On the bases of these observations, the LPSO model structures have been successfully constructed with the aid of first-principles calculations. The main results are summarized as follows.

1. Electron diffraction and TEM-EDX analysis identify the present LPSO phases as 14H-$Mg_{94}Zn_2Y_4$, 14H-$Mg_{83}Zn_{11}Er_6$ and 18R-$Mg_{85}Zn_6Y_9$. Distributions of Zn/RE atoms are restricted at the local *fcc*-stacking layers common for the present long-period stacking series, and the chemical order of $6\times(10\bar{1}0)_{hcp}$ within the close-packed layers is commonly developed for the LPSO phases. Superlattice reflections of the $6\times(10\bar{1}0)_{hcp}$ order distinctly appear for the LPSO phases with relatively Zn/RE-rich contents, although the peaks are significantly diffused. Such superlattice reflections become extremely weak for the LPSO phase with less-Zn/RE contents, suggesting that a degree of the $6\times(10\bar{1}0)_{hcp}$ chemical order is significantly affected by the LPSO compositions. Z-contrast STEM observations of the present LPSO phases confirm that the Zn and/or RE distributions are not limited to the two close-packed layers, as frequently assumed in the previous reports, but indeed extended across the four close-packed layers relevant to the local *fcc*-stacking.

2. Ideal models of the 14H-type and 18R-type LPSO structures are constructed; 14H-$Mg_{35}Zn_3RE_4$ ($Mg_{83.3}Zn_{7.2}RE_{9.5}$) with $P6_3/mcm$, $a$ = 1.11, $c$ = 3.65 nm and 18R-$Mg_{29}Zn_3RE_4$ ($Mg_{80.6}Zn_{8.3}RE_{11.1}$) with $C2/m$, $a$ = 1.11, $b$ = 1.93, $c$ = 1.60 nm, $\beta$ = 76.6° or $P3_212$, $a$ = 1.11, $c$ = 4.69 nm. All the models are composed of the common

*fcc*-stacking units with identical chemical order, whose characteristic feature is well represented by the $L1_2$-type ordered $Zn_6RE_8$ clusters successfully embedded in the local *fcc*-stacking layers so as to form the $6\times(10\bar{1}0)_{hcp}$ superlattice order. Ideal stoichiometry compositions of the LPSO phases are given at the conditions that the $Zn_6RE_8$-cluster density is maximized within the framework of $6\times(10\bar{1}0)_{hcp}$ order.

3. Initial atomic coordinates given as the ideal honeycomb sites are further optimized based on first-principles calculations. After the successful structural relaxations with a reasonable energy gain from the initial configurations, it is found that significant displacements occur particularly for the Zn and RE atom positions to form shorter Zn-RE atomic distances, indicating the strong interactions between Zn and RE atoms within the LPSO structures. Such strong Zn-RE local condensations cause a puckering of the close-packed layers and displacive modulations of the Mg atoms within the close-packed planes. Occurrences of these characteristic modulations in the present LPSO specimens are semi-quantitatively confirmed based on a comparison between the experimental and simulated electron diffraction patterns.

4. With respect to the ideal LPSO stoichiometry derived from the present models, the Mg-Zn-Y LPSO phases practically occur with non-stoichiometry compositions, whose ranges appear to be asymmetrically extended to a less Zn/Y side and bounded by a fixed

Zn/Y ratio around ~1. This suggests that the LPSO phases are able to tolerate a considerable degree of disorder at the Zn and RE sites with statistical co-occupations by Mg, and therefore the degree of order of the LPSO phases depends on occupation conditions of the Zn/RE sites. Local atomic substitution behaviors in the LPSO phases are effectively discussed in terms of formation conditions of the $Zn_6RE_8$ clusters, explaining qualitatively why the non-stoichiometry composition ranges appear to be bounded by a fixed Zn/Y ratio.

**Acknowledgments**

We would be grateful to Professors Kawamura, Itoi, Yamasaki for their help and advices for sample preparations. We would also thank to Professors Ohtani, Higashida and Mr. Ono for stimulating discussions. This work is supported by a Grant-in-Aid for Scientific Research on Priority Areas "Synchro-LPSO Structures'' from the Ministry of Education, Culture, Sports, Science and Technology of Japan (MEXT), and the Kumamoto Prefecture CREATE Project from JST.

**14H-LPSO ($P6_3/mcm$)**

| atom | site | $x$ | $y$ | $z$ |
|---|---|---|---|---|
| Mg1 | 24*l* | 1/6 | 4/6 | 1/28 |
| | | 0.166 | 0.653 | 0.037 |
| Mg2 | 24*l* | 5/6 | 1/6 | 3/28 |
| | | 0.833 | 0.173 | 0.110 |
| Mg3 | 24*l* | 1/6 | 4/6 | 5/28 |
| | | 0.165 | 0.663 | 0.180 |
| Mg4 | 12*k* | 3/6 | 0 | 3/28 |
| | | 0.498 | 0 | 0.107 |
| Mg5 | 12*k* | 5/6 | 0 | 5/28 |
| | | 0.837 | 0 | 0.178 |
| Mg6 | 12*k* | 2/6 | 0 | 5/28 |
| | | 0.332 | 0 | 0.180 |
| Mg7 | 12*j* | 1/6 | 2/6 | 7/28 |
| | | 0.168 | 0.331 | 0.250 |
| Mg8 | 8*h* | 2/6 | 4/6 | 3/28 |
| | | 0.333 | 0.667 | 0.108 |
| Mg9 | 6*g* | 3/6 | 0 | 7/28 |
| | | 0.499 | 0 | 0.250 |
| Mg10 | 4*c* | 2/6 | 4/6 | 7/28 |
| | | 0.333 | 0.667 | 0.250 |
| Mg11 | 2*a* | 0 | 0 | 7/28 |
| | | 0 | 0 | 0.250 |
| Zn | 12*k* | 5/6 | 0 | 1/28 |
| | | 0.768 | 0 | 0.051 |
| Er1 | 12*k* | 2/6 | 0 | 1/28 |
| | | 0.282 | 0 | 0.030 |
| Er2 | 4*e* | 0 | 0 | 3/28 |
| | | 0 | 0 | 0.091 |

**18R-LPSO ($C2/m$)**

| atom | site | $x$ | $y$ | $z$ |
|---|---|---|---|---|
| Mg1 | 8*j* | 2/36 | 1/12 | 1/12 |
| | | 0.060 | 0.082 | 0.082 |
| Mg2 | 8*j* | 2/36 | 3/12 | 1/12 |
| | | 0.054 | 0.249 | 0.083 |
| Mg3 | 8*j* | 2/36 | 5/12 | 1/12 |
| | | 0.055 | 0.418 | 0.085 |
| Mg4 | 8*j* | 11/36 | 2/12 | 1/12 |
| | | 0.307 | 0.168 | 0.082 |
| Mg5 | 8*j* | 11/36 | 4/12 | 1/12 |
| | | 0.306 | 0.334 | 0.080 |
| Mg6 | 8*j* | 3/36 | 2/12 | 3/12 |
| | | 0.085 | 0.166 | 0.248 |
| Mg7 | 8*j* | 3/36 | 4/12 | 3/12 |
| | | 0.088 | 0.329 | 0.243 |
| Mg8 | 8*j* | 12/36 | 1/12 | 3/12 |
| | | 0.328 | 0.084 | 0.244 |
| Mg9 | 8*j* | 12/36 | 3/12 | 3/12 |
| | | 0.333 | 0.251 | 0.250 |
| Mg10 | 8*j* | 30/36 | 1/12 | 3/12 |
| | | 0.841 | 0.087 | 0.242 |
| Mg11 | 8*j* | 7/36 | 2/12 | 5/12 |
| | | 0.189 | 0.173 | 0.415 |
| Mg12 | 8*j* | 34/36 | 1/12 | 5/12 |
| | | 0.959 | 0.083 | 0.414 |
| Mg13 | 8*j* | 34/36 | 3/12 | 5/12 |
| | | 0.938 | 0.244 | 0.414 |
| Mg14 | 4*i* | 11/36 | 0 | 1/12 |
| | | 0.308 | 0 | 0.081 |
| Mg15 | 4*i* | 29/36 | 0 | 1/12 |
| | | 0.802 | 0 | 0.085 |
| Mg16 | 4*i* | 3/36 | 0 | 3/12 |
| | | 0.085 | 0 | 0.250 |
| Zn1 | 8*j* | 16/36 | 1/12 | 5/12 |
| | | 0.424 | 0.116 | 0.381 |
| Zn2 | 4*i* | 25/36 | 0 | 5/12 |
| | | 0.772 | 0 | 0.380 |
| Y1 | 8*j* | 7/36 | 4/12 | 5/12 |
| | | 0.167 | 0.357 | 0.429 |
| Y2 | 4*i* | 21/36 | 0 | 3/12 |
| | | 0.572 | 0 | 0.283 |
| Y3 | 4*i* | 7/36 | 0 | 5/12 |
| | | 0.238 | 0 | 0.430 |

**18R-LPSO ($P3_212$)**

| atom | site | $x$ | $y$ | $z$ | atom | site | $x$ | $y$ | $z$ |
|---|---|---|---|---|---|---|---|---|---|
| Mg1 | 6*c* | 1/18 | 8/18 | 13/36 | Mg22 | 6*c* | 7/18 | 2/18 | 17/36 |
| | | 0.055 | 0.430 | 0.362 | | | 0.392 | 0.115 | 0.473 |
| Mg2 | 6*c* | 4/18 | 5/18 | 13/36 | Mg23 | 6*c* | 7/18 | 11/18 | 17/36 |
| | | 0.235 | 0.289 | 0.362 | | | 0.385 | 0.612 | 0.472 |
| Mg3 | 6*c* | 7/18 | 2/18 | 13/36 | Mg24 | 6*c* | 10/18 | 8/18 | 17/36 |
| | | 0.401 | 0.123 | 0.362 | | | 0.557 | 0.445 | 0.473 |
| Mg4 | 6*c* | 7/18 | 11/18 | 13/36 | Mg25 | 6*c* | 10/18 | 17/18 | 17/36 |
| | | 0.376 | 0.610 | 0.362 | | | 0.552 | 0.942 | 0.473 |
| Mg5 | 6*c* | 10/18 | 17/18 | 13/36 | Mg26 | 6*c* | 13/18 | 5/18 | 17/36 |
| | | 0.542 | 0.943 | 0.362 | | | 0.725 | 0.275 | 0.473 |
| Mg6 | 6*c* | 13/18 | 5/18 | 13/36 | Mg27 | 6*c* | 13/18 | 14/18 | 17/36 |
| | | 0.722 | 0.264 | 0.362 | | | 0.725 | 0.778 | 0.472 |
| Mg7 | 6*c* | 1/18 | 2/18 | 15/36 | Mg28 | 6*c* | 16/18 | 2/18 | 17/36 |
| | | 0.057 | 0.118 | 0.419 | | | 0.889 | 0.109 | 0.473 |
| Mg8 | 6*c* | 1/18 | 11/18 | 15/36 | Mg29 | 6*c* | 16/18 | 11/18 | 17/36 |
| | | 0.062 | 0.609 | 0.419 | | | 0.889 | 0.614 | 0.472 |
| Mg9 | 6*c* | 4/18 | 8/18 | 15/36 | Zn1 | 6*c* | 1/18 | 17/18 | 13/36 |
| | | 0.221 | 0.443 | 0.417 | | | 0.120 | 0.008 | 0.373 |
| Mg10 | 6*c* | 4/18 | 17/18 | 15/36 | Zn2 | 6*c* | 13/18 | 14/18 | 13/36 |
| | | 0.230 | 0.945 | 0.419 | | | 0.657 | 0.777 | 0.373 |
| Mg11 | 6*c* | 7/18 | 5/18 | 15/36 | Zn3 | 6*c* | 16/18 | 11/18 | 13/36 |
| | | 0.391 | 0.279 | 0.417 | | | 0.889 | 0.544 | 0.373 |
| Mg12 | 6*c* | 7/18 | 14/18 | 15/36 | Y1 | 6*c* | 4/18 | 14/18 | 13/36 |
| | | 0.386 | 0.777 | 0.417 | | | 0.175 | 0.777 | 0.357 |
| Mg13 | 6*c* | 10/18 | 2/18 | 15/36 | Y2 | 6*c* | 10/18 | 8/18 | 13/36 |
| | | 0.556 | 0.110 | 0.417 | | | 0.602 | 0.490 | 0.357 |
| Mg14 | 6*c* | 10/18 | 11/18 | 15/36 | Y3 | 6*c* | 16/18 | 2/18 | 13/36 |
| | | 0.547 | 0.604 | 0.419 | | | 0.888 | 0.062 | 0.357 |
| Mg15 | 6*c* | 13/18 | 8/18 | 15/36 | Y4 | 6*c* | 16/18 | 14/18 | 15/36 |
| | | 0.715 | 0.436 | 0.419 | | | 0.889 | 0.777 | 0.406 |
| Mg16 | 6*c* | 13/18 | 17/18 | 15/36 | | | | | |
| | | 0.721 | 0.950 | 0.419 | | | | | |
| Mg17 | 6*c* | 16/18 | 5/18 | 15/36 | | | | | |
| | | 0.889 | 0.275 | 0.416 | | | | | |
| Mg18 | 6*c* | 1/18 | 8/18 | 17/36 | | | | | |
| | | 0.053 | 0.439 | 0.473 | | | | | |
| Mg19 | 6*c* | 1/18 | 17/18 | 17/36 | | | | | |
| | | 0.053 | 0.942 | 0.472 | | | | | |
| Mg20 | 6*c* | 4/18 | 5/18 | 17/36 | | | | | |
| | | 0.225 | 0.278 | 0.473 | | | | | |
| Mg21 | 6*c* | 4/18 | 14/18 | 17/36 | | | | | |
| | | 0.221 | 0.777 | 0.473 | | | | | |

Table 1 Atomic coordinates of model structures of 14H-type and 18R-type LPSO phases; initial positions in fractional numbers (upper row) and energetically optimized positions in decimal fractions (lower row). Initial lattice parameters and atomic sites are derived from *hcp*-Mg structure ($a$ = 0.32, $c$ = 0.52 nm). 14H-type structure (left); $P6_3/mcm$, $a$ = 1.11 (1.11), $c$ = 3.62 (3.65) nm (initial values are in brackets). 18R-type structure; (middle) $C2/m$, $a$ = 1.11 (1.11), $b$ = 1.93 (1.94), $c$ = 1.60 (1.60) nm, $\beta$ = 76.5° (76.6°) or (right) $P3_212$, $a$ = 1.11 (1.11), $c$ = 4.69 (4.69) nm (initial values are in brackets).

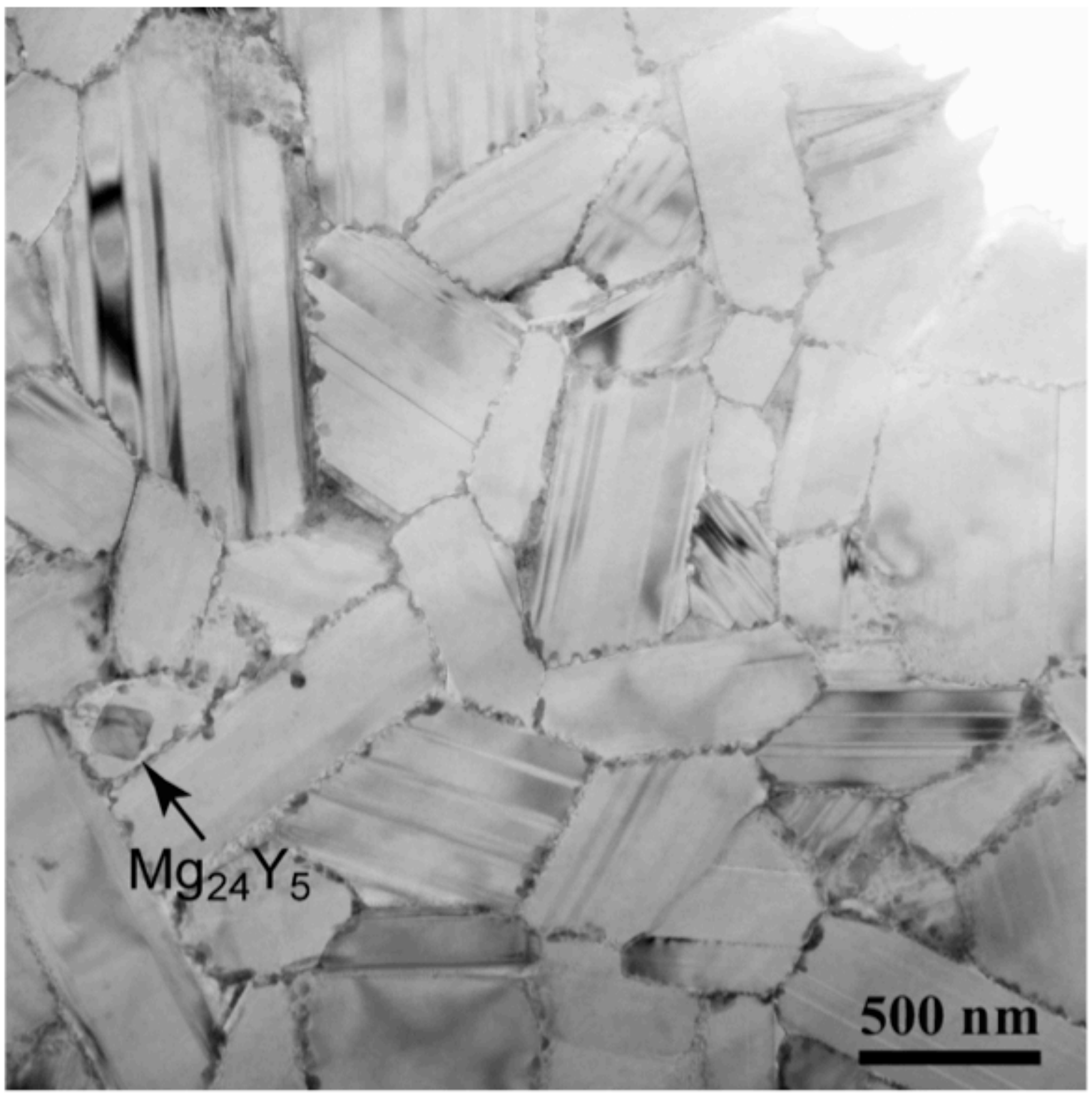

Fig. 1. A bright field TEM image of the $Mg_{85}Zn_6Y_9$ alloy annealed at 673K for 72h. Except for rarely observed $Mg_{24}Y_5$ particles such as indicated by an arrow, the microstructure consists of equiaxed grains of the 18R-type LPSO phase, whose volume fraction is estimated to be more than ~95%.

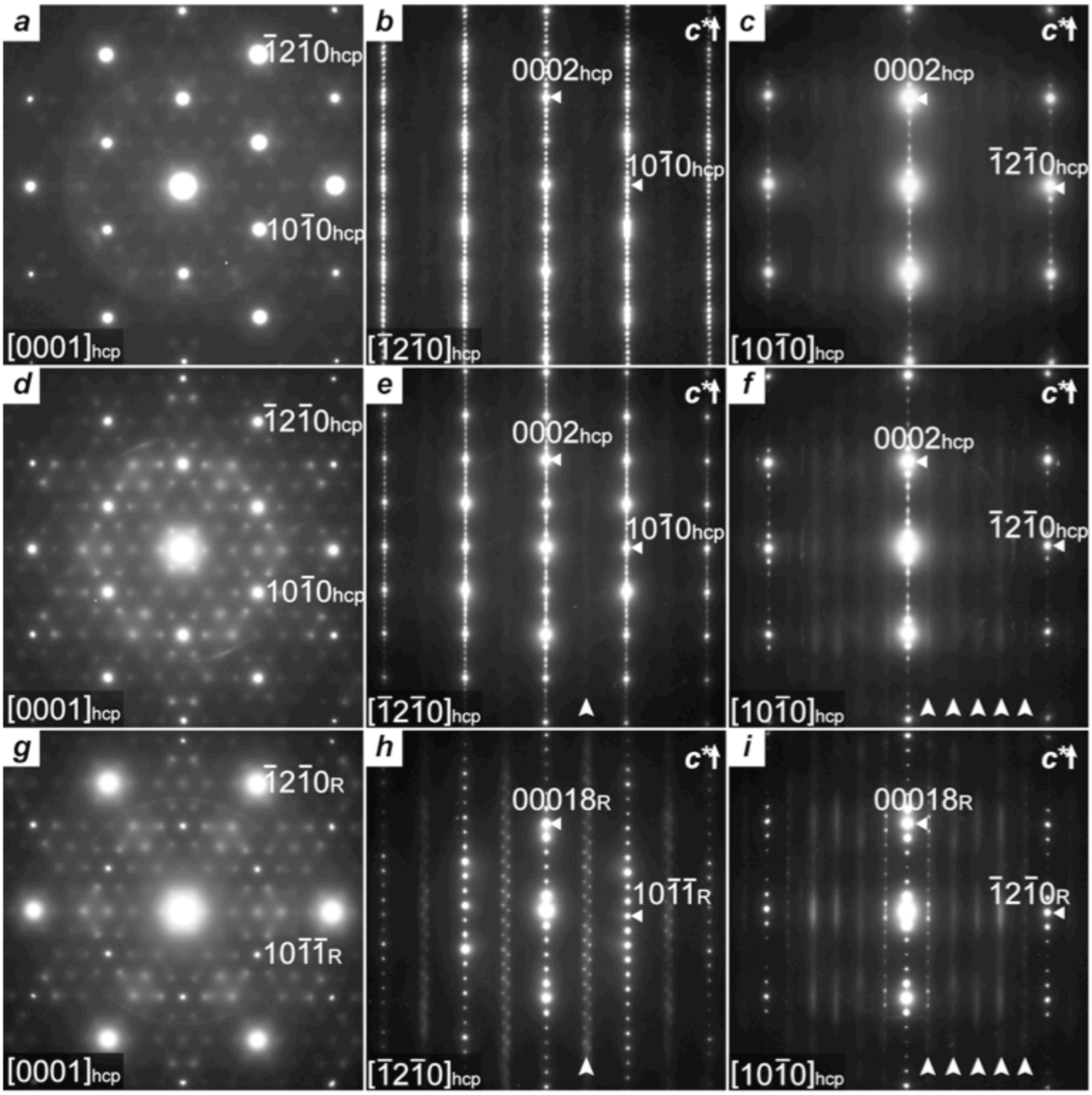

Fig. 2. Electron diffraction patterns obtained from the LPSO phases formed in (a)-(c) $Mg_{97}Zn_1Y_2$, (d)-(f) $Mg_{97}Zn_1Er_2$ and (g)-(i) $Mg_{85}Zn_6Y_9$ alloys. These patterns are taken with the incident beam parallel to (a), (d), (g) $[0001]_{hcp}$, (b), (e), (h) $[\bar{1}2\bar{1}0]_{hcp}$ and (c), (f), (i) $[10\bar{1}0]_{hcp}$ directions, respectively. The indices are based on the *hcp*-Mg structure for (a) – (f) and the fundamental rhombohedral lattice ($R\bar{3}m$, a = 0.322 , c = 4.69 nm) for (g) – (i). Weak reflections significantly diffused and/or streaked along the *c**-axis are indicated by arrowheads in the pattern (e), (f), (h) and (i).

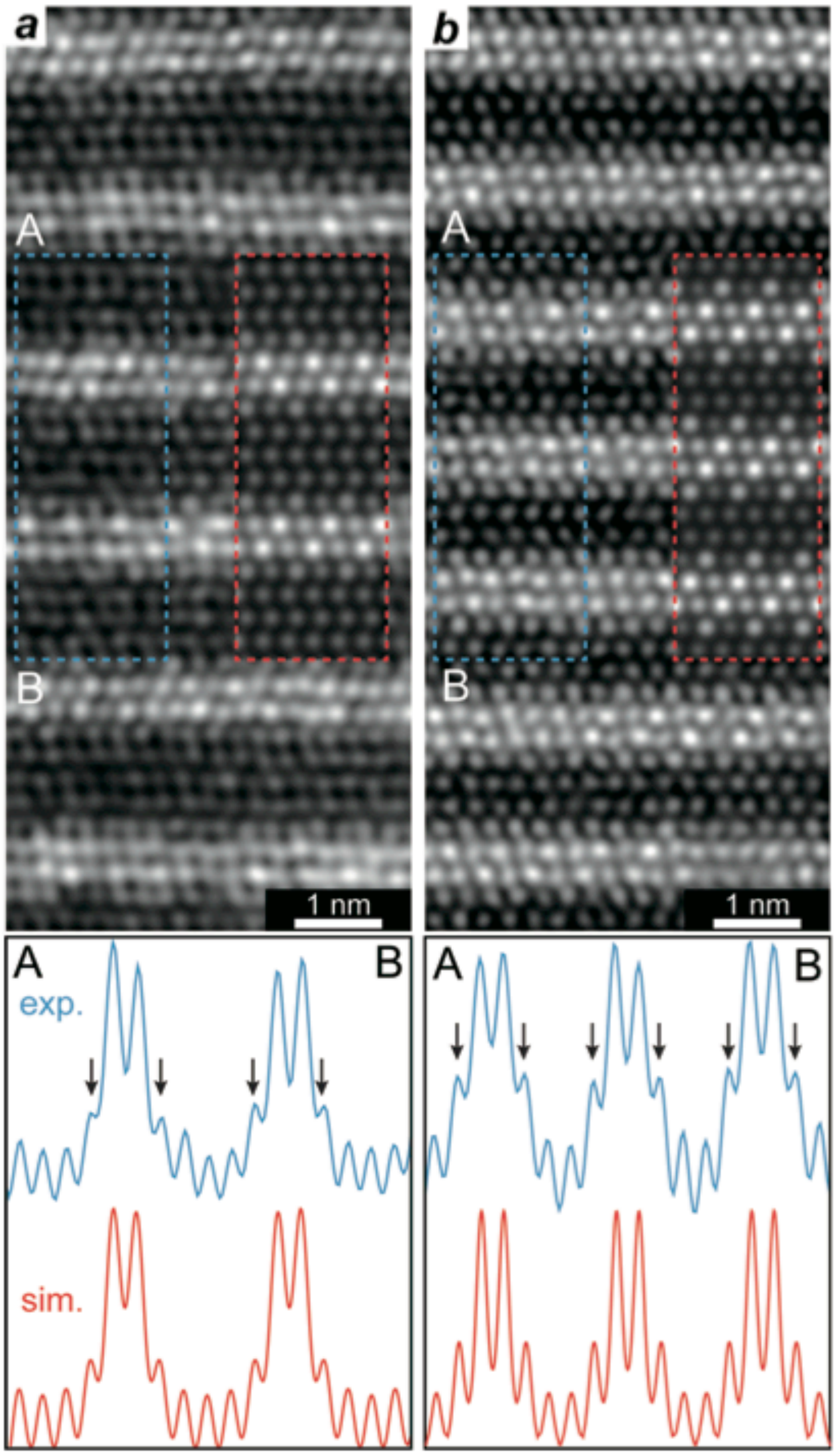

Fig. 3. HAADF-STEM images of (a) 14H-$Mg_{94}Zn_2Y_4$ and (b) 18R-$Mg_{85}Zn_6Y_9$, taken with the diffraction patterns of Figs. 2 (b) and (h), respectively. Intensity profiles along the stacking directions are shown at the each bottom, which are obtained by integrating over the blue-frame area across A-B indicated in (a) and (b). Simulated images based on the present structure models (initial atomic positions denoted in Table 1) are inserted in red-frame area, and the parameters used for the calculations are: (a) a spherical aberration coefficient $C_s$ = 0.5mm, specimen thickness $t \sim$ 15nm, defocus values $\Delta f$ = 35nm, a beam convergence angle $\alpha$ = 12mrad, (b) $C_s$ = 0 mm, $t \sim$ 15 nm, defocus = 0 nm, $\alpha$ = 22mrad. To account for the non-stoichiometry composition of (a), the simulation is made with the occupations at the RE sites 0.33Y/0.67Mg and that at the Zn sites 0.25Zn/0.75Mg, to be approximately ~$Mg_{94}Zn_2Y_4$. The occupations are unity at all the sites for the stoichiometry phase of (b).

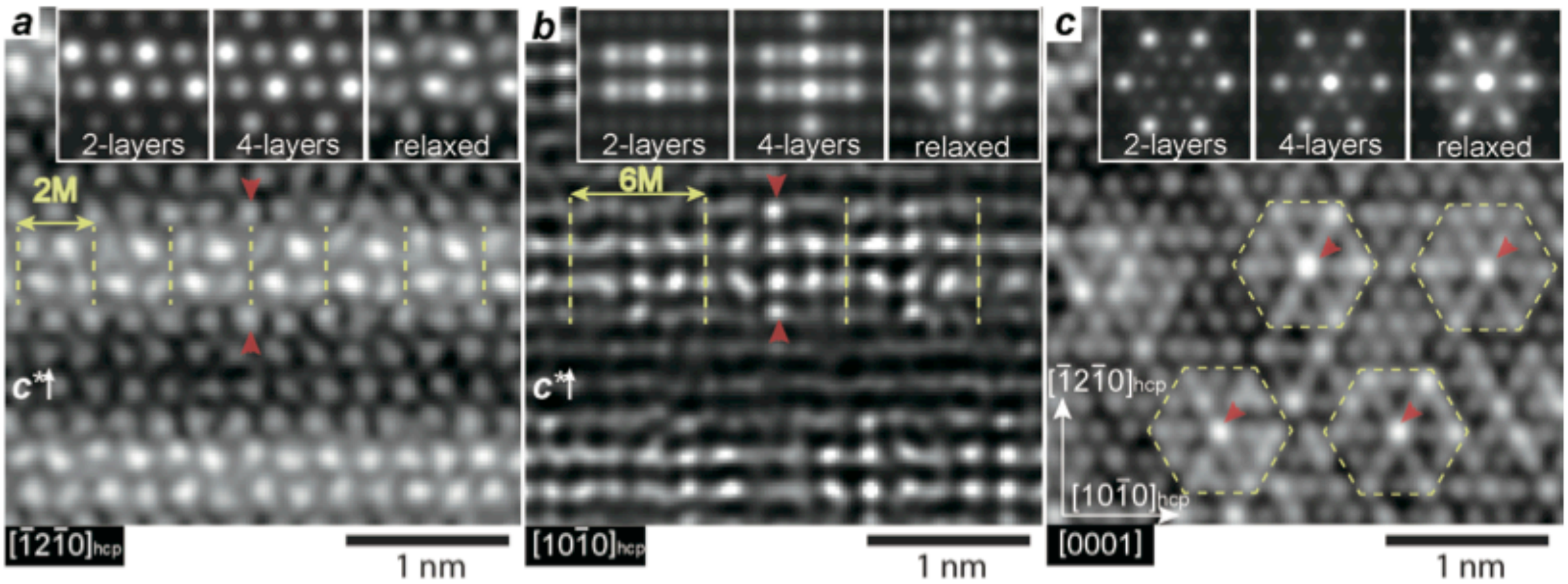

Fig. 4. HAADF-STEM images of the well-ordered LPSO phases taken along the major zone axes; (a) $[\bar{1}2\bar{1}0]_{hcp}$, (b) $[10\bar{1}0]_{hcp}$ and (c) $[0001]_{hcp}$, each of which simulated images based on the present models (Table 1) are inserted at the top. The mages were obtained from (a), (b) 18R-$Mg_{85}Zn_6Y_9$ and (c) 14H-$Mg_{83}Zn_{11}Er_6$. Simulations with initial atomic positions are denoted by '2-layers' and '4-layers', which mean whether or not the simulations include the relevant RE sites indicated by red arrowheads in (a) - (c) (i.e., RE atoms are located only at the two close-packed layers or extended over the four close-packed layers). Simulated images denoted as 'relaxed' are based on the energetically-relaxed atomic positions. Parameters for simulations are same as those described for Fig. 3(b).

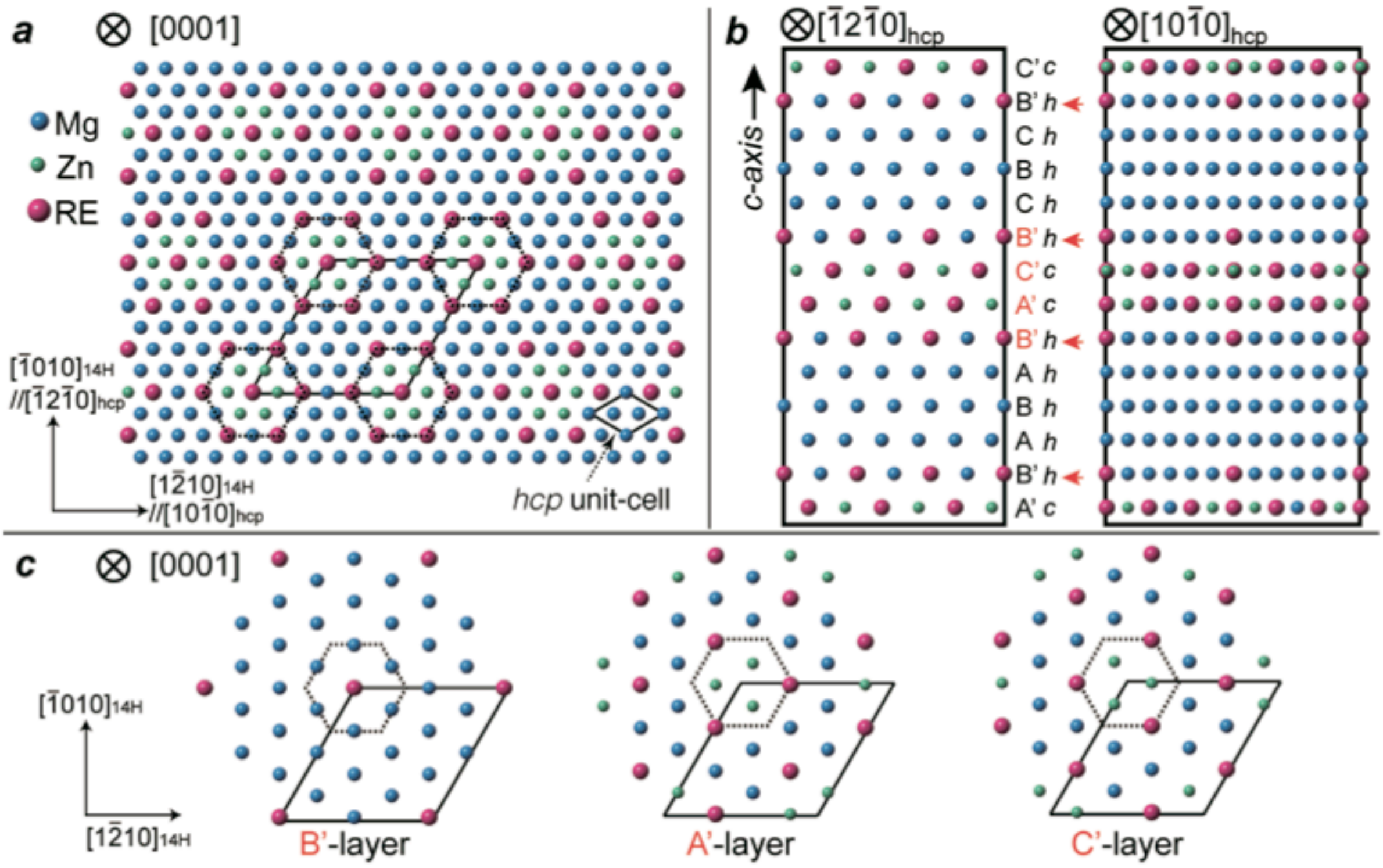

Fig. 5. Structure model of 14H-$Mg_{35}Zn_3RE_4$, shown by (a), (b) projections along the major zone axes and (c) layer-by-layer representations. Unit cell of 14H-LPSO is indicated by the rhombus in (a). In (b), stacking sequences are denoted by ABC together with the Jagodzinski notation, *h* and *c*, which represent the local stacking environment [12]. Zn/RE site distributions within B', A' and C' close-packed layers are shown in (c).

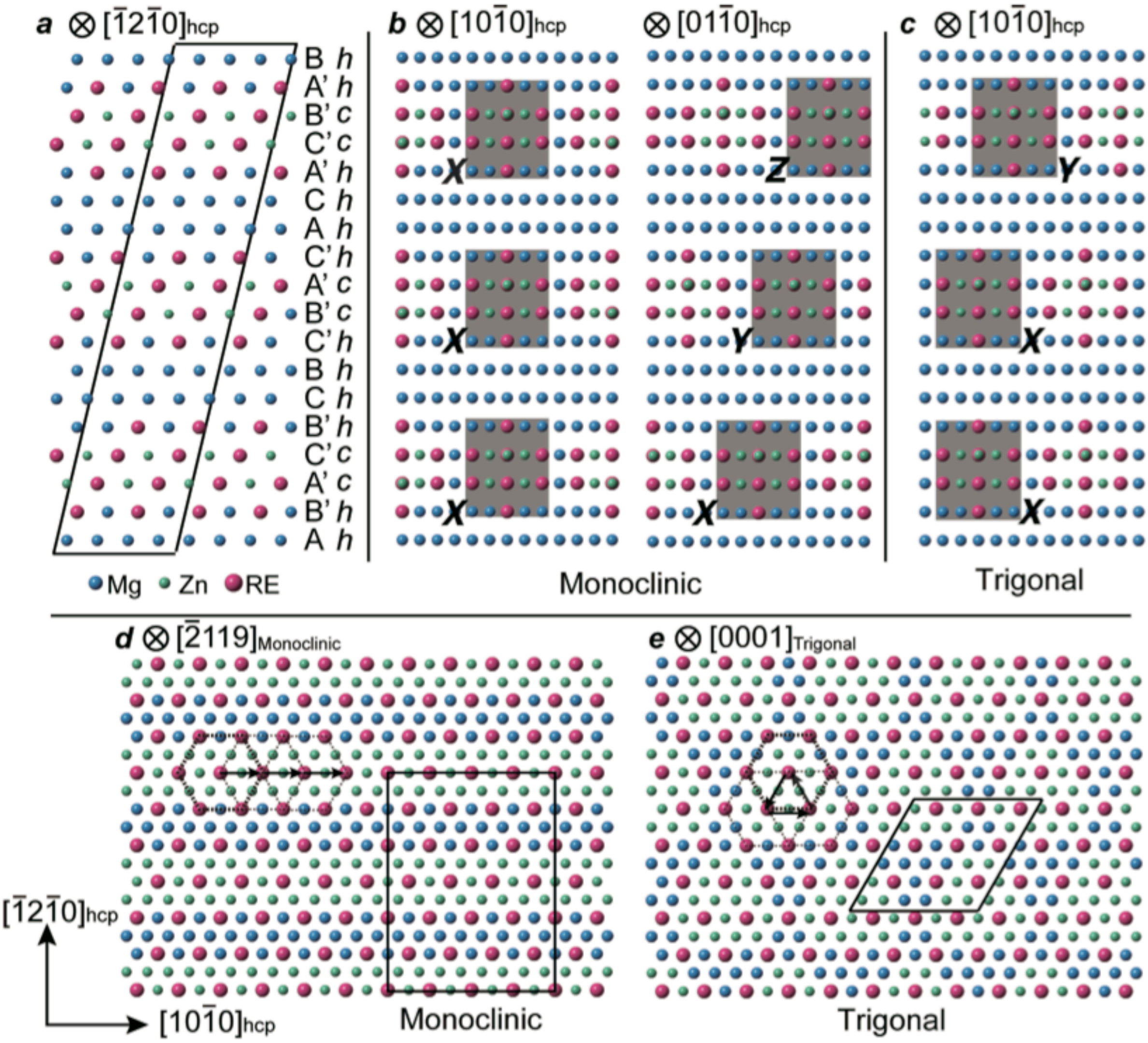

Fig. 6. Structure models of 18R-$Mg_{29}Zn_3RE_4$ with $C2/m$ (monoclinic) and $P3_212$ (trigonal), shown by projections along the major zone axes. (a) Both models appear to be identical Zn/RE distributions in $[\bar{1}2\bar{1}0]_{hcp}$ projection, in which stacking sequences are denoted by ABC together with the Jagodzinski notation, *h* and *c*, which represent the local stacking environment [12]. (b) - (d) Model differences are seen in the projections along (b), (d) $[10\bar{1}0]_{hcp}$ and (d), (e) $[0001]_{hcp}$. $L1_2$-type $Zn_6RE_8$-cluster positions are exemplified by (b), (d) gray-square and (d), (e) dashed-hexagon.

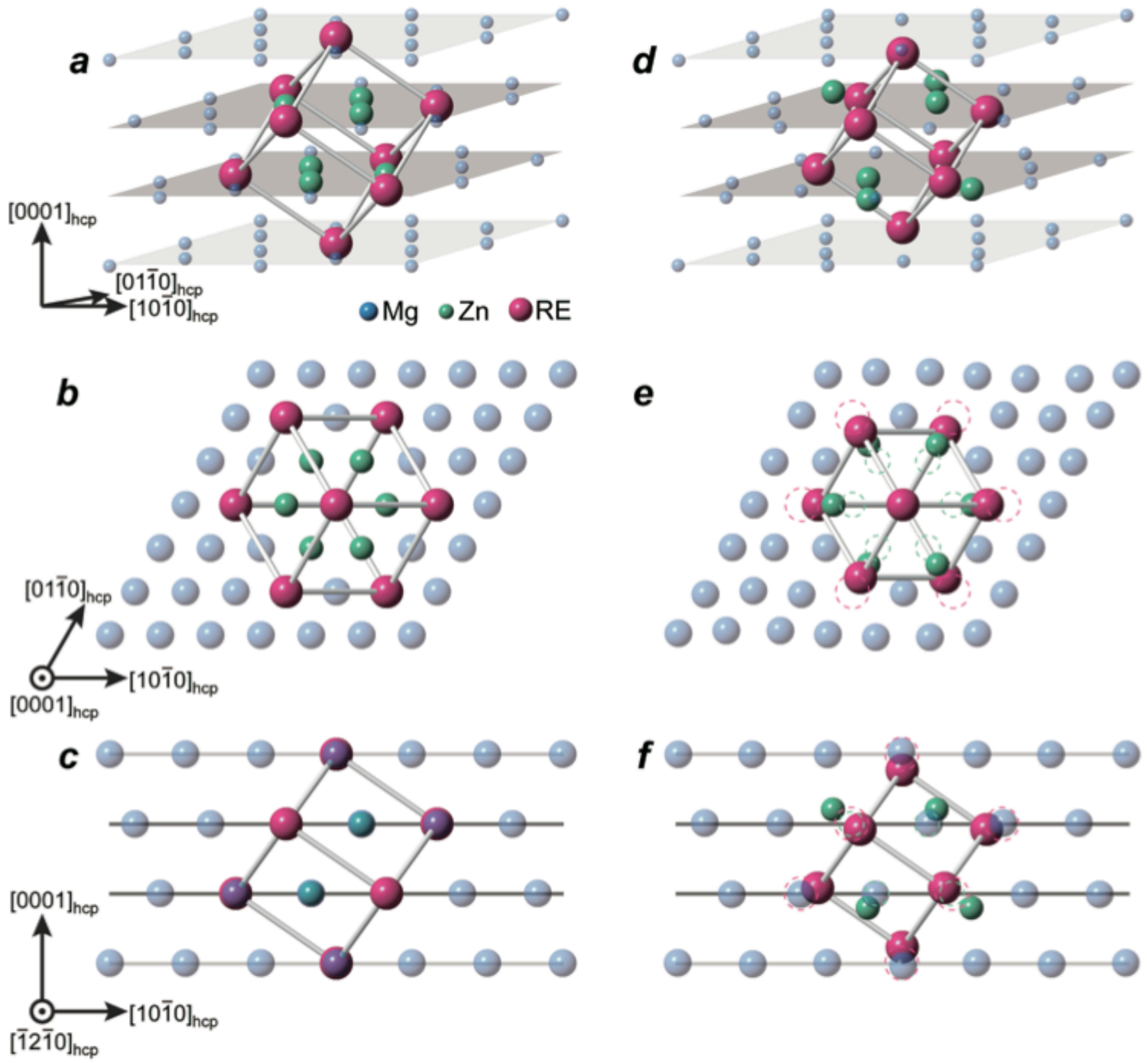

Fig. 7. Local structures around the $Zn_6RE_8$ cluster embedded in *fcc*-stacking layers in the LPSO; (a)-(c) initial configuration and (d)-(e) energetically optimized configuration. Structures are shown (a), (d) schematic views, (b), (e) $[0001]_{hcp}$ projection and (c), (f) $[\bar{1}2\bar{1}0]_{hcp}$ projection. For computations, the cut off energy was chosen at 340 eV, with *k*-mesh = 3 × 3 × 3. Structural models contain 168 atoms for 14H-type and 144 atoms (monoclinic) and 216 atoms (trigonal) for 18R-type LPSO structures, which have been relaxed within space group symmetry by allowing the unit cell volume to expand/contract.

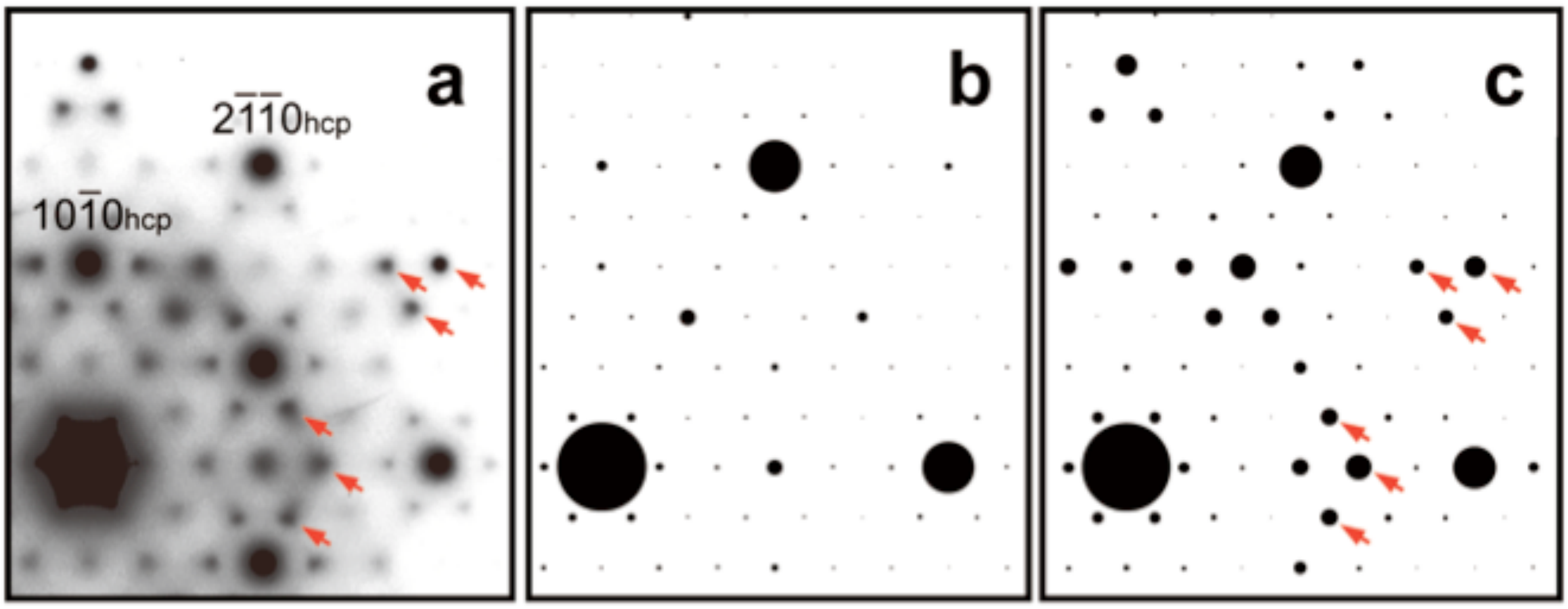

Fig. 8. Electron diffraction patterns with $[0001]_{hcp}$ incidence of the 14H-type LPSO. (a) Experimental pattern, enlargement of Fig. 2 (d). (b), (c) Calculated kinematical patterns (structure factors) of 14H-$Mg_{35}Zn_3RE_4$ structure with (b) initial and (c) optimized atom positions.

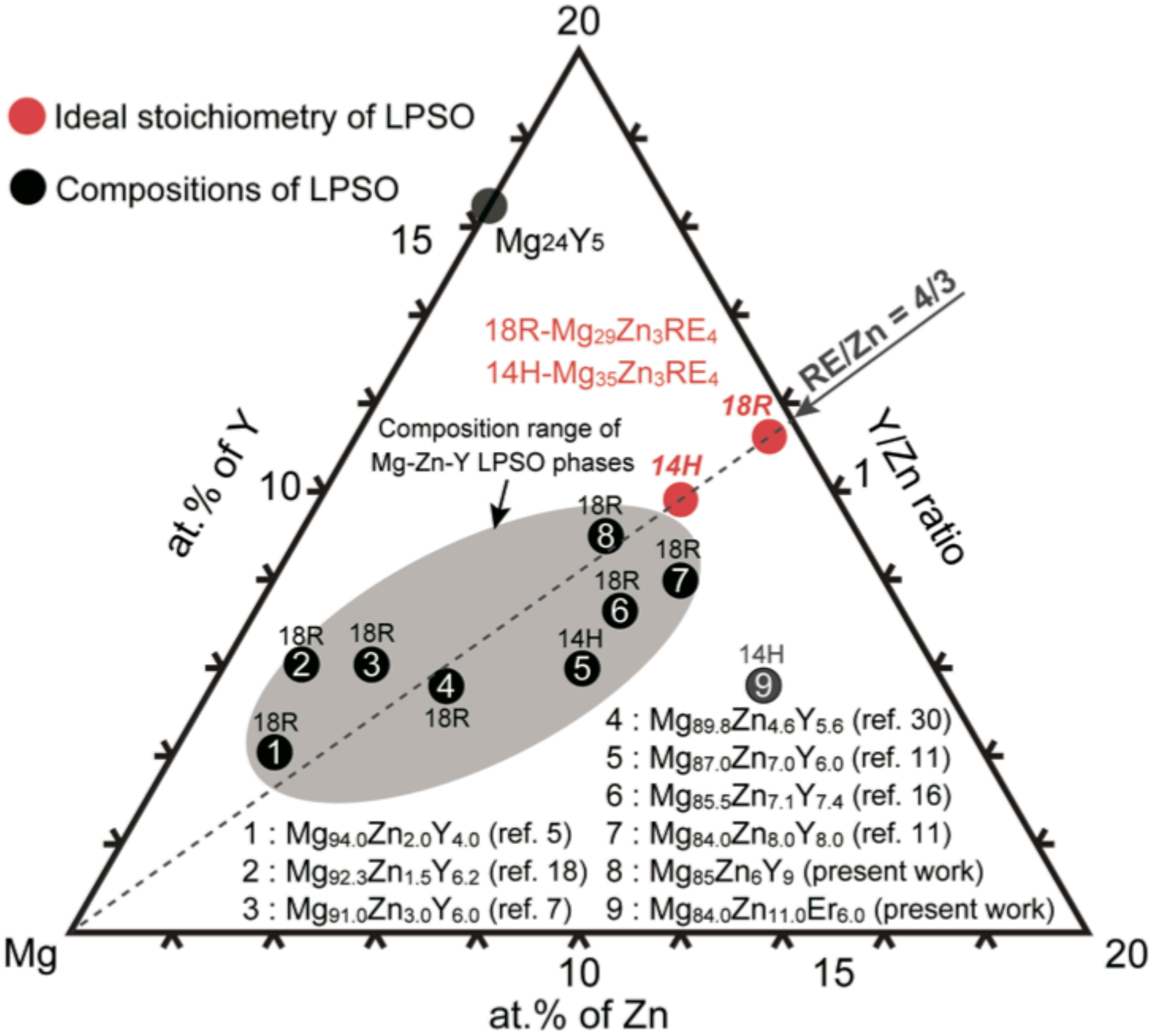

Fig. 9. Schematic quasi-isothermal section of Mg-Zn-Y ternary phase diagram. Experimentally determined compositions of the Mg-Zn-Y LPSO phases, annealed at temperatures 573K ~ 793K, are plotted together with ideal stoichiometry compositions of the present LPSO models (red). The 14H-$Mg_{83}Zn_{11}Er_6$ is also plotted as No. 9 for comparison.